# Spatially homogeneous conformal Stäckel spacetimes of type (3.1)

*Evgeniy Konstantinovich Osetrin* *

*Konstantin Evgenievich Osetrin* **

*Altair Evgenievich Filippov* ***

*Tomsk State Pedagogical University*

**Abstract**

In this work, we obtain all spatially homogeneous space-time models related to the intersection of the set of Stäckel spaces of type (2.1) and the set of conformally Stäckel spaces of type (3.1). These models allow a complete separation of variables both in the Hamilton-Jacobi equation for massive test particles and in the eikonal equation for radiation. The models obtained in this work relate to wave-like space-time models. For the found models, solutions of the Einstein equations with the cosmological constant and radiation are obtained. For the obtained solutions, the eikonal equation and the equations of motion of massive test particles in the Hamilton-Jacobi form are integrated.

**Keywords:** theory of gravity, exact solutions, group of motions, homogeneous spaces, Killing fields, gravitational waves, Petrov's classification, Bianchi classification.

* Email: evgeny.osetrin@gmail.com
** Email: osetrin@tspu.edu.ru
*** Email: altair@tspu.edu.ru

# Spatially homogeneous conformal Stäckel spacetimes of type (3.1)

*Konstantin Osetrin, Evgeniy Osetrin, Altair Filippov*
Tomsk State Pedagogical University

Hamilton-Jacobi equation of a test particle in a gravitational field has the form:

$$g^{ij}\frac{\partial S}{\partial x^i}\frac{\partial S}{\partial x^j} = m^2, \quad i,j = 0\ldots 3, \tag{1}$$

Where $S = S(x^i)$ - the function of the test particle action, $g^{ij}$ - spacetime metric.

In order not to be confused, we note that we will also use not only a cursive (capital) letter $S$ to indicate the function of the test particle action but also a lowercase cursive letter $s$ to indicate the space-time interval.

Spaces that allow complete separation of variables in the Hamilton-Jacobi equation for the test particles (1) are called Stäckel spaces named after Paul Stäckel see [1, 2].

The equation for radiation in a gravitational field has the form:

$$\tilde{g}^{ij}\frac{\partial \Psi}{\partial x^i}\frac{\partial \Psi}{\partial x^j} = 0, \tag{2}$$

Where $\Psi = \Psi(x^i)$ - the eikonal function, $\tilde{g}^{ij}$ - spacetime metric.

Spaces admitting a complete separation of variables in the eikonal equation (2) are called conformal Stäckel spaces.

Considering Stäckel spaces of type (2.1). According to the general theory of Stäckel spaces developed by V.N. Shapovalov [3,4], these spaces allow the existence of a privileged coordinate system (CS), for which the eikonal equation and the Hamilton-Jacobi equation for a test particle allow integration by the method of complete separation of variables. Stäckel spaces are known to be determined by the so-called "complete set" of Killing fields, consisting of Killing vectors and Killing tensors of the second rank [5]. Stäckel spaces of type (2.1) admit two Killing commuting vectors in the complete set, therefore, in a privileged CS (where the separation of variables is allowed), the metric of the Stäckel space of type (2.1) can be written so that it depends only on two variables - $x^0$ and $x^1$:

$$g^{ij} = \frac{1}{\Delta}\begin{pmatrix} 1 & 0 & 0 & 0 \\ 0 & 0 & f_1(x^1) & 1 \\ 0 & f_1(x^1) & A(x^0, x^1) & b_0(x^0) \\ 0 & 1 & b_0(x^0) & c_0(x^0) \end{pmatrix}, \tag{3}$$

Where $A(x^0, x^1) = a_0(x^0) + a_1(x^1)$, a conformal multiplier $\Delta$ is in case of conformal Stäckel spaces an arbitrary function of all four variables, and in the case of Stäckel spaces of type (2.1) it has the form $\Delta = t_0(x^0) + t_1(x^1)$. Variables on which the metric in a privileged CS does not depend are called ignored (cyclic).

If the metric of the Stäckel space of type (2.1) in the privileged CS (3) depends only on one variable $x^1$, and the conformal factor still has the form $\Delta = t_0(x^0) + t_1(x^1)$, we obtain spaces from a subset of conformal Stäckel spaces of type (3.1).

For Stäckel spaces, the equation (1) allows integration by the method of complete separation of variables and the action function $S$ can be written in a privileged coordinate system in an additively "divided" form (4). In the same coordinate system, it allows a complete separation of variables in the eikonal equation, while the metric $\tilde{g}^{ij}$ will be different from the $g^{ij}$ with the presence of an arbitrary conformal factor $\Delta$.

The ability to integrate geodesic equations in the Stäckel spaces makes it possible to obtain precisely integrable models in metric theories of gravity [6] - [16], including modified gravity theories [17] - [19], and gravitational-wave astronomy.

For Stäckel spaces of type (2.1) (with metric (3) in the privileged CS) for the function of test particles action, we have (ignored variable metric enter linearly):

$$S = \phi_0(x^0) + \phi_1(x^1) + px^2 + qx^3, \; p, q, r - \text{const}. \tag{4}$$

And, the functions $\phi_0(x^0)$ and $\phi_1(x^1)$ are the solutions of ordinary differential equations:

$$\dot{\phi}_0^2 = m^2 t_0(x^0) - p^2 a_0(x^0) - 2pq b_0(x^0) - q^2 c_0(x^0) - r, \tag{5}$$

$$2\dot{\phi}_1 \left( pf_1(x^1) + q \right) = m^2 t_1(x^1) - p^2 a_1(x^1) + r, \tag{6}$$

Where $m$, $p$, $q$, $r$ - constant parameters of the test particles. The dot above means ordinary differentiation.

Thus, in the space-time models under consideration, it is possible to integrate the quadratures of the test particle motion equation in the Hamilton-Jacobi form and find the eikonal function, i.e. geodesic propagation of radiation in a gravitational field. In the case of the eikonal equation, the vector $l_k = \partial \Psi / \partial x^k$ sets the radiation wave vector, and the equations $\Psi(x^i) = \text{const}$ sets the radiation wave surface (wavefront).

**Classes of the spatially homogeneous Stäckel spaces (SS) models (2.1)**

When isolating spatially homogeneous models from the family of Stäckel spaces of type (2.1), we assume that the number of pairwise commuting Killing vectors of the considered models remain equal to two so that the metric in a privileged CS depends on two non-ignored variables, and commuting Killing vectors $X_0$ and $X_1$ from the complete set of SS (2.1) can be represented as:

$$X_0^i = (0,0,0,1), \qquad X_1^i = (0,0,1,0). \tag{7}$$

Note that the vector $X_0$ is isotropic, and the vector $X_1$ is spatially similar.

The additional two Killing vectors providing spatial homogeneity of the models are marked as follows:

$$X_2 = \xi^i \partial_i, \qquad X_3 = \eta^j \partial_j \tag{8}$$

Models of spatially homogeneous Stäckel spaces of type (2.1), allowing in a privileged CS the dependence of the metric on one of the ignored variables only through the conformal factor, belong to conformal Stäckel spaces of type (3.1). It is these models that will be considered in this paper. We will use the name "model of class B of spatially homogeneous SS (2.1)" to denote them, in contrast to the models of class A, the set of which does not intersect with the set of conformal Stäckel spaces of type (3.1).

In total, we have identified two non-reducible models of spatially homogeneous SS (2.1) class B, which exhaust all models of the class in question.

For each model under consideration, integration of the Einstein equations with the cosmological constant $\Lambda$ and the energy-momentum tensor of pure radiation with the energy density $\varepsilon$ and wave vector $l^k$ will be carried out below:

$$R_{ij} - \frac{1}{2} R g_{ij} = \Lambda g_{ij} + \varepsilon l_i l_j, \qquad l^k l_k = 0. \tag{9}$$

For models that correspond to Einstein's equations (9), the eikonal equation and the equation of test particles motion in the Hamilton-Jacobi form will also be integrated below.

**Model B.1 spatially homogeneous SS (2.1)**

For this class of the spatially homogeneous SS(2.1) we have:

$$f_1 = 0, \quad a_0 = 0, \quad A = a_1 = \frac{1}{x^{1^2} - \alpha^2}\left(\frac{x^1 - \alpha}{x^1 + \alpha}\right)^\beta, \quad \alpha\beta \neq 0,$$

$$b_0 = 0, \quad c_0 = 0, \quad t_1 = 0, \quad \Delta = t_0 = 1/(x^0)^2. \tag{10}$$

$$g^{ij} = x^{0^2}\begin{pmatrix} 1 & 0 & 0 & 0 \\ 0 & 0 & 0 & 1 \\ 0 & 0 & (x^1-\alpha)^{-1+\beta}(x^1+\alpha)^{-1-\beta} & 0 \\ 0 & 1 & 0 & 0 \end{pmatrix}, \quad \xi^i = \begin{bmatrix} x^0 \\ 0 \\ x^2 \\ 2x^3 \end{bmatrix}, \quad \eta^i = \begin{bmatrix} x^0 x^1 \\ x^{1^2} - \alpha^2 \\ \alpha\beta x^2 \\ -x^{0^2}/2 \end{bmatrix}, \tag{11}$$

The space-time interval has the form:

$$ds^2 = \frac{1}{x^{0^2}}\left[dx^{0^2} + 2dx^1 dx^3 + (x^1-\alpha)^{1-\beta}(x^1+\alpha)^{1+\beta} dx^{2^2}\right], \tag{12}$$

Where $\alpha$ and $\beta$ - constant model parameters, $x^1$ - wave (isotropic) variable.

Killing vector commutators of the model B.1 have the form:

$[X_0, X_1] = 0, \ [X_0, X_2] = 2X_0, \ [X_0, X_3] = 0, \ [X_1, X_2] = X_1, \ [X_1, X_3] = \alpha\beta X_1, \ [X_2, X_3] = 0.$

When $\beta = 0$ this space admits an additional Killing commuting vector and degenerates into a Stäckel space type (3.1).

The condition of the spatial homogeneity of the model imposes restrictions to the region of acceptable coordinate values: $\left((x^{0^2} + 2x^1 x^3)^2 - 4\alpha^2 x^{3^2}\right)(x^{1^2} - \alpha^2) < 0$. From the form of the metric determinant, it follows that $|x^1| > |\alpha|$, then the spatial homogeneity of the model additionally requires that the condition for acceptable coordinate values be met: $|x^{0^2} + 2x^1 x^3| < 4|\alpha x^3|$.

The scalar curvature of the model is constant and negative $R = -12$, and the components of the Weyl tensor are proportional to the following expressions $C_{ijkl} \sim \alpha(\beta^2 - 1)$.

Considering the integration of Einstein's equations (9) for the metric (11). We get the following solution:

$$\Lambda = 3, \quad l_0 = l_2 = l_3 = 0, \quad \varepsilon l_1(x^1)^2 = \frac{\alpha^2(1-\beta^2)}{(x^{1^2} - \alpha^2)^2}, \quad |\beta| \leq 1, \quad \alpha\beta \neq 0. \tag{13}$$

Thus we obtain a spatially homogeneous wave-like Universe with the interval (12), where the variable $x^1$ - wavelike, with a cosmological constant $\Lambda$, filled with radiation with the energy density $\varepsilon$ and a wave radiation vector $l^k = (0,0,0,l_1 x^{0^2})$, $l^3 = x^{0^2} l_1(x^1)$.

The obtained spatially homogeneous space-time model is type III according to the Bianchi classification and type N according to the Petrov's classification.

With the parameter value $\beta = 1$ the model degenerates and becomes a vacuum and conformally flat.

**Integration of the Hamilton-Jacobi equation and the eikonal equation for model B.1**

We integrate equations (5) - (6) for the metric (11) and obtain the explicit form of the function $S$ for the test particle action (4) (when $q \neq 0$):

$$S = m \ln x^0 \pm \left[ \sqrt{m^2 + rx^{0^2}} - m \ln \left( m + \sqrt{m^2 + rx^{0^2}} \right) \right]$$

$$- \frac{2\alpha\beta r x^1 + p^2 \left[ (x^1 - \alpha)/(x^1 + \alpha) \right]^\beta}{4q\alpha\beta} + px^2 + qx^3 + F(p,q,r). \tag{14}$$

Here $p,q,r$ - the independent constant of motion of test particles, set by the initial conditions.

The eikonal function for the metric (11) (when $q \neq 0$) will take the form:

$$\Psi = rx^0 - \frac{2\alpha\beta r^2 x^1 + p^2 \left[ (x^1 - \alpha)/(x^1 + \alpha) \right]^\beta}{4q\alpha\beta} + px^2 + qx^3 + F(p,q,r). \tag{15}$$

In the particular case when the constant of the test particle motion $q$ goes to zero value, the constants $p$ and $r$ go to zero value as well. Then for the action of the test particle function, we obtain:

$$S = m \ln x^0. \tag{16}$$

**Model B.2 spatially homogeneous SS (2.1)**

For this class of the spatially homogeneous SS(2.1) we have:

$$f_1 = 0, \quad a_0 = 0, \quad A = a_1 = x^{1(2\alpha-1)}, \quad b_0 = 0, \quad c_0 = 0, \quad t_1 = 0, \quad \Delta = t_0 = 1/(x^0)^2, \quad \alpha \neq 0, 1/2. \tag{17}$$

$$g^{ij} = x^{0^2} \begin{pmatrix} 1 & 0 & 0 & 0 \\ 0 & 0 & 0 & 1 \\ 0 & 0 & x^{1(2\alpha-1)} & 0 \\ 0 & 1 & 0 & 0 \end{pmatrix}, \quad \xi^i = \begin{bmatrix} x^0 \\ 0 \\ x^2 \\ 2x^3 \end{bmatrix}, \quad \eta^i = \begin{bmatrix} x^0/2 \\ x^1 \\ \alpha x^2 \\ 0 \end{bmatrix}, \tag{18}$$

The space-time interval has the form:

$$ds^2 = \frac{1}{x^{0^2}} \left( dx^{0^2} + 2dx^1 dx^3 + x^{1(1-2\alpha)} dx^{2^2} \right), \tag{19}$$

Where $\alpha$ is a constant model parameter, $x^1$ - wave variable.

Killing vector commutators of the model B.2 have the form:
$[X_0, X_1] = 0, \; [X_0, X_2] = 2X_0, \; [X_0, X_3] = 0, \; [X_1, X_2] = X_1, \; [X_1, X_3] = \alpha X_1, \; [X_2, X_3] = 0.$

With $\alpha = 0$ this space admits an additional Killing commuting vector and degenerates into a Stäckel space type (3.1), that is why we consider that $\alpha \neq 0$.

The condition of the spatial homogeneity of the model imposes restrictions to the region of acceptable coordinate values: $x^1 x^3 (x^{0^2} + 2x^1 x^3) < 0$. Since it follows from the form of the metric determinant that $x^1 > 0$, then the condition of spatial homogeneity of the model requires a restriction to the acceptable range of values of variables with the form $x^3(x^{0^2} + 2x^1 x^3) < 0$, whence it follows that $x^3 < 0$ and $x^{0^2} > 2|x^1 x^3|$.

The scalar curvature of the model is constant and negative $R = -12$, and the components of the Weyl tensor are proportional to the following expressions $C_{ijkl} \sim (4\alpha^2 - 1)$

Considering the integration of Einstein's equations (9) for the metric (18). We get the following solution:

$$\Lambda = 3, \qquad l_0 = l_2 = l_3 = 0, \qquad \varepsilon l_1^2 = \frac{1 - 4\alpha^2}{4x^{1^2}}, \qquad |\alpha| \leq 1/2. \tag{20}$$

Thus we obtain a spatially homogeneous wave-like Universe (where the variable $x^1$ - wavelike) with a cosmological constant $\Lambda$ filled with radiation with energy density $\varepsilon$ and a wave radiation vector $l^k = (0, 0, 0, l_1 x^{0^2})$, $l^3 = x^{0^2} l_1$.

The obtained spatially homogeneous space-time model is of type III according to the Bianchi classification and type N according to the Petrov classification.

With parameter value $\alpha = 1/2$ the model B.2 degenerates - the metric in a privileged CS depends only on one variable $x^0$, the model becomes a vacuum and conformally flat.

**Integration of the Hamilton-Jacobi equation and the eikonal equation for the model B.2**

We integrate equations (5) - (6) for the metric (18) and obtain the explicit form of the function $S$ for the test particle action (4) (when $q \neq 0$):

$$S = m \ln x^0 + \sqrt{m^2 - rx^{0^2}} - m \ln\left(m + \sqrt{m^2 - rx^{0^2}}\right) + \frac{2\alpha r x^1 - p^2 (x^1)^{2\alpha}}{4\alpha q} + px^2 + qx^3 + F(p, q, r). \tag{21}$$

Here $p$, $q$, $r$ - the independent constant motion of test particles, set by the initial conditions.

The solution of the eikonal equation when $q \neq 0$ will take a simple form:

$$\Psi = rx^0 - \frac{1}{2q}\left(r^2 x^1 + \frac{p^2}{2\alpha} x^{1^{2\alpha}}\right) + px^2 + qx^3 + F(p, q, r), \tag{22}$$

Where $p$, $q$, $r$ - independent constants defined by the initial conditions.

In a degenerate case, with the constant $q$ turning to zero, the constants $p$, $r$ also vanish and we obtain the test particle action:

$$S = m \ln x^0. \tag{23}$$

**Conclusion**

In this work, we found spatially homogeneous space-time models that allow the existence of privileged coordinate systems in which the Hamilton-Jacobi equation of the test particles allows integration by the method of complete separation of variables according to type (2.1). In the work, only the models of class B were considered which relate also to conformal Stäckel spaces of type (3.1) and they allow integration of the eikonal equation.

The considered models relate to the wave-like space-time models, which metrics in a privileged coordinate system (where the separation of variables is allowed) depend on the wave (isotropic) variable.

Two models of the class were obtained. The obtained spatially homogeneous space-time models are of type III according to the Bianchi classification and type N according to the Petrov classification.

For the obtained space-time models, exact solutions of the Einstein equations with a cosmological constant and the energy-momentum tensor of pure radiation were found - aperiodic wave-like exact solutions of the Einstein equations for a spatially homogeneous Universe with a nonzero cosmological constant were obtained.

For the obtained spatially homogeneous exact solutions of the Einstein equations with radiation and the cosmological constant, an explicit form of the full integrals for the eikonal function and the action of massive test particles were found.

The study was carried out with the financial support of the Russian Federal Property Fund for research project No. 18-31-00040.

## List of references